\documentclass[9pts,conference,hidelinks]{IEEEtran}
\IEEEoverridecommandlockouts
\usepackage{cite}
\usepackage{amsmath,amssymb,amsfonts}
\usepackage{graphicx}
\usepackage{textcomp}
\usepackage{balance}
\usepackage{xcolor}
\usepackage{algorithm}
\usepackage[noend]{algpseudocode}
\usepackage{mathtools}
\usepackage{multirow}
\usepackage{enumitem}
\usepackage{caption}
\usepackage{subcaption}
\usepackage{calc}
\usepackage{comment}
\usepackage{etoolbox}
\usepackage{booktabs}
\usepackage{adjustbox}
\usepackage{listings}
\usepackage{xcolor}
\usepackage[colorinlistoftodos]{todonotes}

\newcommand\dac{\textcolor{black}{AutoMM}}
\newcommand{\jz}[1]{\textcolor{red}{#1}}
\newcommand{\cb}[1]{\textcolor{blue}{#1}} 

\def\baselinestretch{0.93}

\usepackage{url}

\usepackage{breakurl}

\definecolor{codegreen}{rgb}{0,0.6,0}
\definecolor{codegray}{rgb}{0.5,0.5,0.5}
\definecolor{codepurple}{rgb}{0.58,0,0.82}
\definecolor{backcolour}{rgb}{0.95,0.95,0.92}

\lstdefinestyle{mystyle}{
    backgroundcolor=\color{backcolour},   
    commentstyle=\color{codegreen},
    keywordstyle=\color{magenta},
    numberstyle=\tiny\color{codegray},
    stringstyle=\color{codepurple},
    basicstyle=\ttfamily\footnotesize,
    breakatwhitespace=false,         
    breaklines=true,                 
    captionpos=b,                    
    keepspaces=true,                 
    numbers=left,                    
    numbersep=5pt,                  
    showspaces=false,                
    showstringspaces=false,
    showtabs=false,                  
    tabsize=2,
    moredelim=**[is][\color{red}]{@}{@}1
}
\begin{document}

\title{\LARGE 
{\dac}: Energy-Efficient Multi-Data-Type Matrix Multiply Design on Heterogeneous Programmable System-on-Chip
\\
\small 60th DAC 2023: \footnotesize{High Performance, Low Power Matrix Multiply Design on ACAP: from Architecture, Design Challenges and DSE Perspectives}
\thanks{This paper is published in 60th DAC 2023. Please cite our DAC paper in your reference: High Performance, Low Power Matrix Multiply Design on ACAP: from Architecture, Design Challenges and DSE Perspectives.}
\vspace{-8pt}
}
\author{\IEEEauthorblockN{Jinming Zhuang}
\IEEEauthorblockA{\textit{University of Pittsburgh} \\
jinming.zhuang@pitt.edu}
\and
\IEEEauthorblockN{Zhuoping Yang}
\IEEEauthorblockA{\textit{University of Pittsburgh} \\
zhuoping.yang@pitt.edu}
\and
\IEEEauthorblockN{Peipei Zhou}
\IEEEauthorblockA{\textit{University of Pittsburgh} \\
peipei.zhou@pitt.edu}
}

\makeatletter
\patchcmd{\@maketitle}
  {\addvspace{0.5\baselineskip}\egroup}
  {\addvspace{-1.0\baselineskip}\egroup}
  {}
  {}
\makeatother

\maketitle



\begin{abstract}
As the increasing complexity of Neural Network(NN) models leads to high demands for computation, AMD introduces a heterogeneous programmable system-on-chip (SoC), i.e., Versal ACAP architectures featured with programmable logic(PL), CPUs, and dedicated AI engines (AIE) ASICs which has a theoretical throughput up to 6.4 TFLOPs for FP32, 25.6 TOPs for INT16 and 102.4 TOPs for INT8. However, the higher level of complexity makes it non-trivial to achieve the theoretical performance even for well-studied applications like matrix-matrix multiply. In this paper, we  provide {\dac}, an automatic white-box framework that can systematically generate the design for MM accelerators on Versal which achieves 3.7 TFLOPs, 7.5 TOPs, and 28.2 TOPs for FP32, INT16, and INT8 data type respectively. 
Our designs are tested on board and achieve gains of 7.20x (FP32), 3.26x (INT16), 6.23x (INT8) energy efficiency than AMD U250 FPGA, 2.32x (FP32) than Nvidia Jetson TX2 GPU, 1.06x (FP32), 1.70x (INT8) than Nvidia A100 GPU.
\end{abstract}

\begin{IEEEkeywords}
heterogeneous system-on-chip, Versal ACAP, matrix multiply, support for multiple data types.
\end{IEEEkeywords}
\section{Introduction}
\label{sec:intro}
With the end of the Dennard voltage scaling law,
domain-specific accelerators, e.g. FPGAs, TPUs, and GPUs, became a mainstream trend to improve performance while maintaining power efficiency~\cite{dally2020domain}. 
To keep up the pace of high computation demand, AMD proposes the Versal architecture which is a heterogeneous programmable system-on-chip featuring the dedicated AI Engine (AIE) ASIC, programmable logic (FPGA), and software ARM cores to provide high throughput while maintaining flexibility.
As shown in Table~\ref{tbl:comparison_energy_efficiency}, we use the on board result of the MM application to demonstrate the energy efficiency between last and current generation FPGAs and GPUs. Comparing the 16nm U250 FPGA with Nvidia Jetson TX2 GPU, Jetson TX2 achieves 3.11x energy efficiency since the bit level reconfiguration of prior FPGAs leads to more power consumption. The 7nm VCK190 enables both bit-level hardware customization on the PL side and byte-level customization on the dedicated AIE array. 
Due to the AIE array, our proposed design on VCK190, i.e., AutoMM, achieves 1.06x energy efficiency compared with Nvidia A100 GPU with the same technology node.

However, designing energy-efficient accelerators on Versal platforms can be very challenging due to the inconsistency between the high throughput provided by the AIE array and the relatively low off-chip bandwidth.
We collect the theoretical performance and off-chip bandwidth of two 16nm and 7nm GPUs and FPGAs under FP32 data type in Table~\ref{tbl:comparison_preformance_bandwidth}. 
The required computation-to-communication (CTC) ratio refers to the minimum data reuse rate that can sustain the theoretical throughput under the provided off-chip bandwidth. 
While VCK190 provides 6400 GFLOPs throughput, it only equips with one DDR4-DIMM external memory with 25.6 GB/s bandwidth meaning at least 250 operations per byte are needed to sustain the peak performance which is 13.10x, 17.01x, and 19.8x more severe compared with 16nm U250 FPGA, 16nm Jetson TX2 GPU and 7nm A100 GPU respectively. 
Therefore, huge challenges caused by the significant gap between performance and off-chip bandwidth on Versal platforms should be addressed to achieve high performance and energy-efficient designs. 
With such contradictory results from energy efficiency and required CTC ratio, one key question arises: \textit{How can we design more energy-efficient MM accelerator designs to make full use of the gigantic computation resources
under limited communication bandwidth?}
To answer this, we identify the design challenges at different levels and show the detailed design methodologies to tackle them:
\begin{table}
\Huge
\caption{Performance, power, and energy efficiency comparisons among FPGAs and GPUs when the data type is FP32.}
\vspace{-10pt}
\label{tbl:comparison_energy_efficiency}
    \begin{center}
    \begin{adjustbox}{width=1\columnwidth,center}
\begin{tabular}{ c | c | c | c | c c }

 \toprule
  \multirow{2}{*}{\textbf{\textcolor{black}{Fabrication}}}  
  &  \multirow{2}{*}{\textbf{\textcolor{black}{Board Name \& Framework}}} 
  &  \textbf{\textcolor{black}{Performance}}  
  & \textbf{\textcolor{black}{Power}} 
  & \multicolumn{2}{c}{\textbf{\textcolor{black}{Energy Efficiency}}} 
\\ & & (GFLOPS) & (Watt) & (GFLOP/J) & (Ratio) \\
    \midrule
    \multirow{2}{*}{16 nm} 
                        & \multicolumn{1}{c}{AMD U250~\cite{u250_web}, AutoSA~\cite{wang2021autosa}}
                        & \multicolumn{1}{c}{858}
                        & \multicolumn{1}{c}{96.20} 
                        & \multicolumn{1}{c}{8.92 }
                        & \multicolumn{1}{c}{1.00x} 
                        \\
                        & \multicolumn{1}{c}{Nvidia Jetson TX2~\cite{tx2_web}, cuBLAS~\cite{cublas}} 
                        & \multicolumn{1}{c}{560}
                        & \multicolumn{1}{c}{20.20} 
                        & \multicolumn{1}{c}{27.72}
                        & \multicolumn{1}{c}{3.11x} 
                        \\
                        \hline
    \multirow{2}{*}{7 nm}
                        
                        & \multicolumn{1}{c}{\cb{AMD VCK190~\cite{vck190_web}, \textbf{This work}}} 
                        & \multicolumn{1}{c}{\cb{3,745}}
                        & \multicolumn{1}{c}{\cb{58.34}}
                        & \multicolumn{1}{c}{\cb{64.18}}
                        & \multicolumn{1}{c}{\cb{7.20x}}
                        \\
                        & \multicolumn{1}{c}{Nvidia A100~\cite{a100_web}, cuBLAS~\cite{cublas}} 
                        & \multicolumn{1}{c}{15,016}
                        & \multicolumn{1}{c}{248.20} 
                        & \multicolumn{1}{c}{60.50}
                        & \multicolumn{1}{c}{6.78x} 
                        \\

    \bottomrule
\end{tabular}
\end{adjustbox}
    \end{center}
    \vspace{-10pt}
\end{table}
\begin{table}
\Huge
\caption{Theoretical performance, off-chip bandwidth and require CTC ratio comparisons among FPGAs and GPUs of two generations when the data type is FP32.}
\vspace{-10pt}
\label{tbl:comparison_preformance_bandwidth}
    \begin{center}
    \begin{adjustbox}{width=1\columnwidth,center}
\begin{tabular}{ c | c | c | c | c c }

 \toprule
  \multirow{2}{*}{\textbf{\textcolor{black}{Fabrication}}}  
  &  \multirow{2}{*}{\textbf{\textcolor{black}{Board Name}}}
  &  \textbf{\textcolor{black}{Performance}}  
  & \textbf{\textcolor{black}{Off-Chip BW}} 
  & \multicolumn{2}{c}{\textbf{\textcolor{black}{Required CTC Ratio}}} \\ & & (GFLOPS) & (GB/s) & (GFLOP/Byte) & (Ratio)
\\
    \midrule
    \multirow{2}{*}{16 nm} 
                        & \multicolumn{1}{c}{AMD U250~\cite{u250_web}} 
                        & \multicolumn{1}{c}{1,470}
                        & \multicolumn{1}{c}{77} 
                        & \multicolumn{1}{c}{19.09}
                        & \multicolumn{1}{c}{1.00x} 
                        \\
                        & \multicolumn{1}{c}{Nvidia Jetson TX2~\cite{tx2_web}} 
                        & \multicolumn{1}{c}{750}
                        & \multicolumn{1}{c}{51.2} 
                        & \multicolumn{1}{c}{14.65}
                        & \multicolumn{1}{c}{0.77x} 
                        \\
                        \hline
    \multirow{2}{*}{7 nm}
                        & \multicolumn{1}{c}{AMD VCK190~\cite{vck190_web}} 
                        & \multicolumn{1}{c}{6,400}
                        & \multicolumn{1}{c}{25.6} 
                        & \multicolumn{1}{c}{250}
                        & \multicolumn{1}{c}{13.10x} 
                        \\
                        & \multicolumn{1}{c}{Nvidia A100~\cite{a100_web}} 
                        & \multicolumn{1}{c}{19,500}
                        & \multicolumn{1}{c}{1555} 
                        & \multicolumn{1}{c}{12.54}
                        & \multicolumn{1}{c}{0.66x}
                        \\

    \bottomrule
\end{tabular}
\end{adjustbox}
    \end{center}
    \vspace{-20pt}
\end{table}

\begin{figure*}
    \vspace{-5pt}
    \centering
    \includegraphics[width=0.85\linewidth]{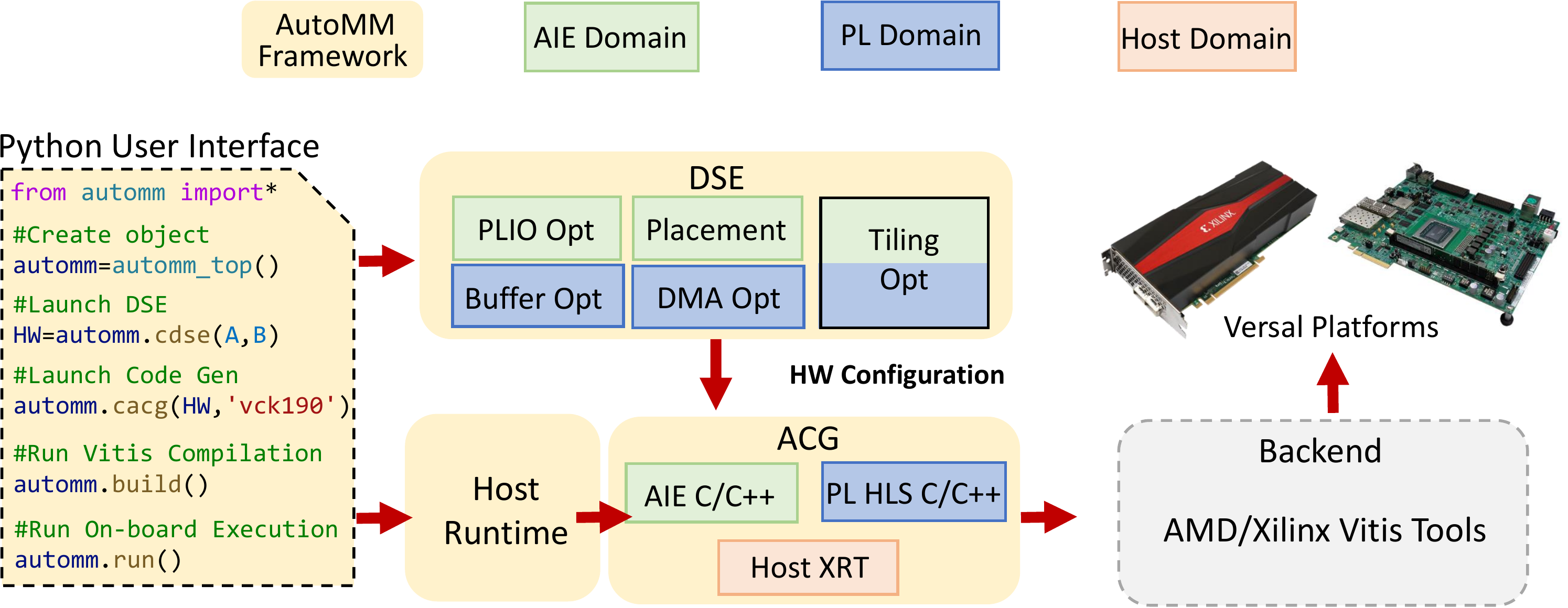}
    \caption{AutoMM Compilation Framework and Python User Interface.}
    \label{fig:framework}
\end{figure*}

\begin{itemize}[leftmargin=*]
\item \textbf{High Efficiency Single AIE Design}: 
To achieve high efficiency in single AIE computation, we propose the optimized coding style in Section~\ref{sec:single_aie} that makes full use of the 7-way VLIW capability to achieve back-to-back issued MAC intrinsic execution.

\item \textbf{IO Reused and Routing Optimized AIE Array Design}: 
We efficiently utilize the limited I/O ports between PL$\leftrightarrow$AIEs by combining broadcast with packet-switch connections to scale out and maintain the computation efficiency to tens and hundreds of AIEs. 
In addition, to alleviate the routing congestion in the AIE array, we explore a broadcast factor for the data transfer from PLIO to AIEs.

\item \textbf{PL$\leftrightarrow$AIEs Bubble-free Pipelining Data Transfer}: 
To amortize the bandwidth gap between limited off-chip bandwidth and the high bandwidth requirement from AIEs, we make full use of the on-chip storage to increase data reuse on PL. Bubble-free pipelining data transfer algorithm is proposed and implemented in the dedicated data mover on PL to feed the data between PL$\leftrightarrow$AIEs producing a non-stall AIE execution pipeline.  

\item We compare the energy efficiency of our design with 16nm U250 FPGA, 16 nm Nvidia Jetson TX2 and 7nm A100 GPU under FP32, INT16, and INT8 data types for MM, NCF, and MLP applications. Our on broad experiment shows that we achieve 3.7 TFLOPs, 7.5 TOPs, and 28.2 TOPs throughput for FP32, INT16, and INT8 on MM. 
Compared with A100 GPU on end-to-end applications, we achieve 0.96x and 1.16x energy efficiency gains on NCF~\cite{he2017neural} and MLP~\cite{wang2019benchmarking}.

\item \textbf{\underline{Auto}matic \underline{MM} Accelerator Design Framework on Versal}:
While AMD provides users a black-box IP DPU~\cite{xilinxdpu} for INT8 neural network (NN) applications, we are among the first ones to provide an open-source white-box framework as shown in Fig~\ref{fig:framework}, i.e., AutoMM, to automatically generate MM accelerator designs for different data types on Versal ACAP. We provide the AutoMM Python APIs to generate the source code for the accelerators.
AutoMM is integrated into CHARM~\cite{fpga2023charm} framework: \textbf{\emph{\url{https://github.com/arc-research-lab/CHARM}}}.
\end{itemize}




\section{Related Work}
\label{sec:2_re_work}

In this section, we discuss the related work of artificial intelligence accelerators on different architectures including FPGAs, GPUs, and dataflow architectures. 

\textbf{FPGA acceleration.} 
Moss~et~al.~\cite{moss2018customizable} propose a customizable hardware template with a fixed systolic array architecture to process matrix multiplication workloads on FPGA.
AutoSA~\cite{wang2021autosa} generates systolic array designs from user-specified matrix sizes by exploring different mapping strategies and implementing them on FPGA.
FBLAS~\cite{de2020fblas} proposes an open-source HLS implementation of the BLAS library for FPGAs.
CHARM (FPGA23~\cite{fpga2023charm}) proposes an open-source design framework of FP32 matrix-multiply-based applications on Versal ACAP (advanced compute acceleration platform).

\textbf{Dataflow architectures.} Eyeriss~\cite{chen2016eyeriss} propose a tiled architecture with a 2D array of PEs and a shared global buffer to process the GEMM operations in NN applications. TPUs~\cite{jouppi2021ten} leverages systolic array architecture to schedule the byte-level computations and data movements in GEMM processing. 

In computation, Versal ACAP is capable of both bit-level computation customization on FPGA and byte-level computation customization as most of the aforementioned dataflow architectures and coarse-grained reconfigurable architecture~\cite{wijerathne2022panorama,cong2014fully,fpga2023charm} support. 
In memory architecture, FPGA and aforementioned dataflow architectures use scratchpad memory, while GPUs~\cite{volkov2008benchmarking} use cache hierarchy to ease the data movement programming.
Versal ACAP also adopts scratchpad memory and therefore, it needs specific control for data movement.  
Specifically, as for on-chip communication, the aforementioned dataflow architectures adopt certain bus-based network-on-chip (NoC) or systolic arrays for the data movements between buffers and computation processing elements.
However, since there is heterogeneity between FPGA and AIE array on Versal ACAP, new challenges including how to efficiently leverage the DMAs and I/Os between FPGA \& AIE arrays and switch-box based AXI stream~(AXIS) within AIE arrays on Versal ACAP need to be solved, and these challenges are addressed in this paper.
\section{Versal Architecture Overview}

\label{sec:3_Versal_arch}
\begin{figure}
\centering
\includegraphics[width=0.9\linewidth]{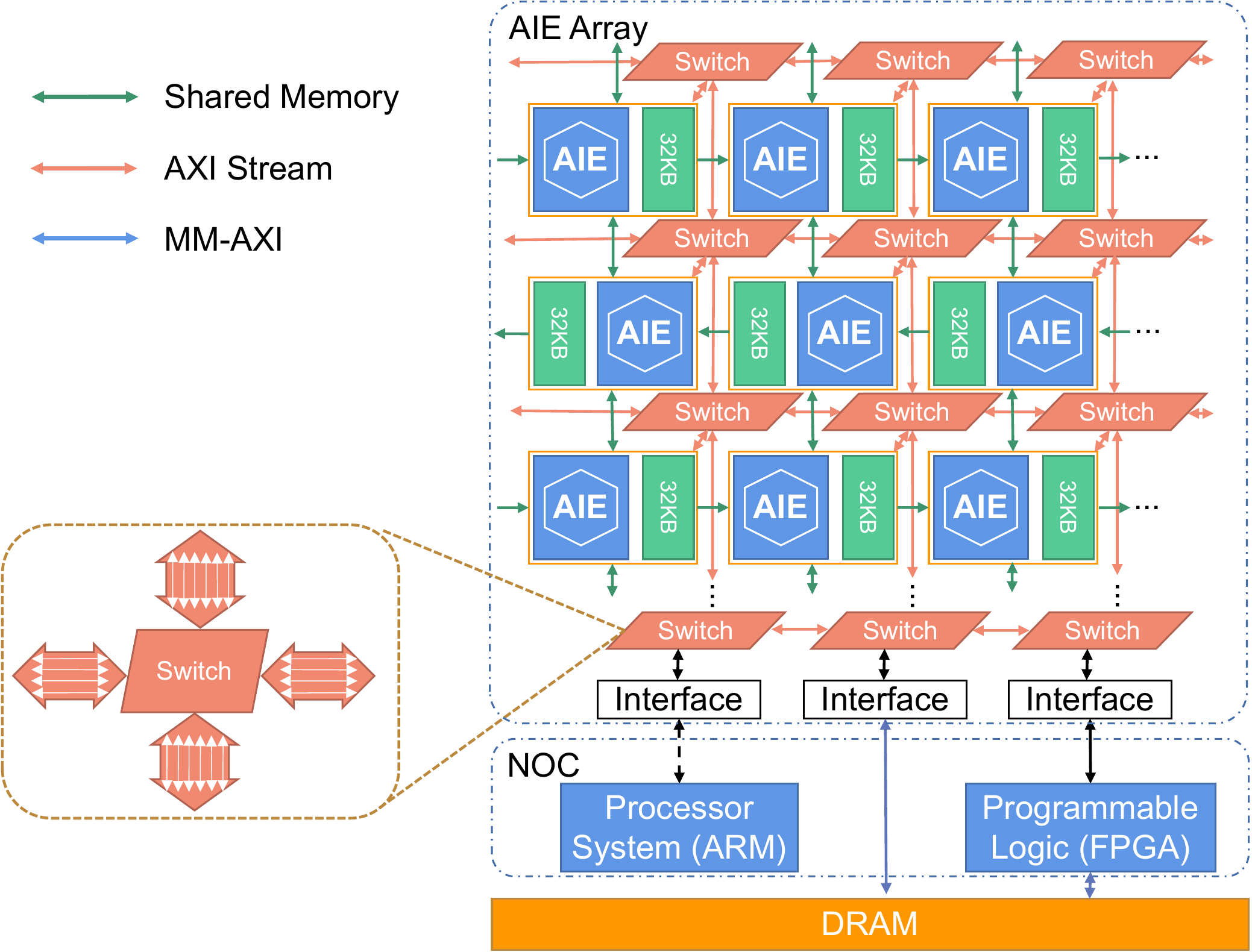}
\caption{Versal ACAP architecture.}
\label{fig:versal-acap-sys-arch}
\vspace{-18pt}
\end{figure}

In this section, we summarize the system architecture of the heterogeneous SoC research platform, AMD Versal VCK190 evaluation kit.  With the AMD XCVC1902 Adaptive Compute Acceleration Platform~(ACAP) chip on the board, VCK190 is featured with a comprehensive set of various hardware as shown in Fig~\ref{fig:versal-acap-sys-arch}.

VCK190 has a wide range of architectures built in, including an array of 400~VLIW processors, called the AI engine array~(AIE array), ARM processors called the Processor System~(PS), and the FPGA Programmable Logic~(PL).  These hardware components could communicate with each other through the NOC or on-chip AXIS.

Inside the AIE array, an AIE core can communicate with another core in two approaches.  Each AIE core shares its local memory with its neighbors for communication.  On the other hand, the cores are connected to an AXI stream mesh through AXIS switches.  The AXIS switches can be reconfigured in such a way that there could be either (1)~a circuit-switched path with dedicated ports for each communication, or (2)~a packet-switched network with a target identifier attached to reuse the paths for multiple communications.  Each AIE core has two input and two output connections from/to the switch.  Each switch has six output ports to its northern neighbor, thus six input ports from its southern neighbor.  For the rest of the directions, the switch has four I/O ports with its neighbor. There are 39 AXIS interface tiles between the AIE array and the PL. The interface crosses the clock domain of the PL and the AIE and automatically converts the rates. The AIE side of the interface has eight 32-bit input and six 32-bit output channels at 1~GHz, supporting up to 256~Gbps input and 192~Gbps output.  The PL side has eight 64-bit input channels and six 64-bit output channels. 

Each AI engine is a 7-way very long instruction word (VLIW) supported vector processor including two loads (from local memory to register), two moves (update vector registers), one store (from register to local memory), one vector operation (2D-SIMD) and one scalar operation instructions. It owns 2Kb vector registers, 3Kb accumulation registers, and 32~KB of data memory located either on the west or the east of the core alternating between rows. In this case, the AIE can not only access its own memory, but also the memory of the AIE on its north and south, and the opposite side of its own memory.  In total, one AIE can access up to 128~KB memory in total.

\section{Design Methodology}
\label{sec:4_design}

Designing a high performance system-level accelerator leveraging heterogeneous resources can be very challenging. 
In this section, we first illustrate the dataflow, tiling, and mapping strategy of  matrix-matrix multiply (MM). We then describe the detailed programming models and design methodologies of the single AIE, AIE array, and  AIE$\leftrightarrow$PL.
\subsection{Dataflow, Tiling and Mapping Strategy of MM}
\begin{figure}
\centering
\includegraphics[width=0.9\linewidth]{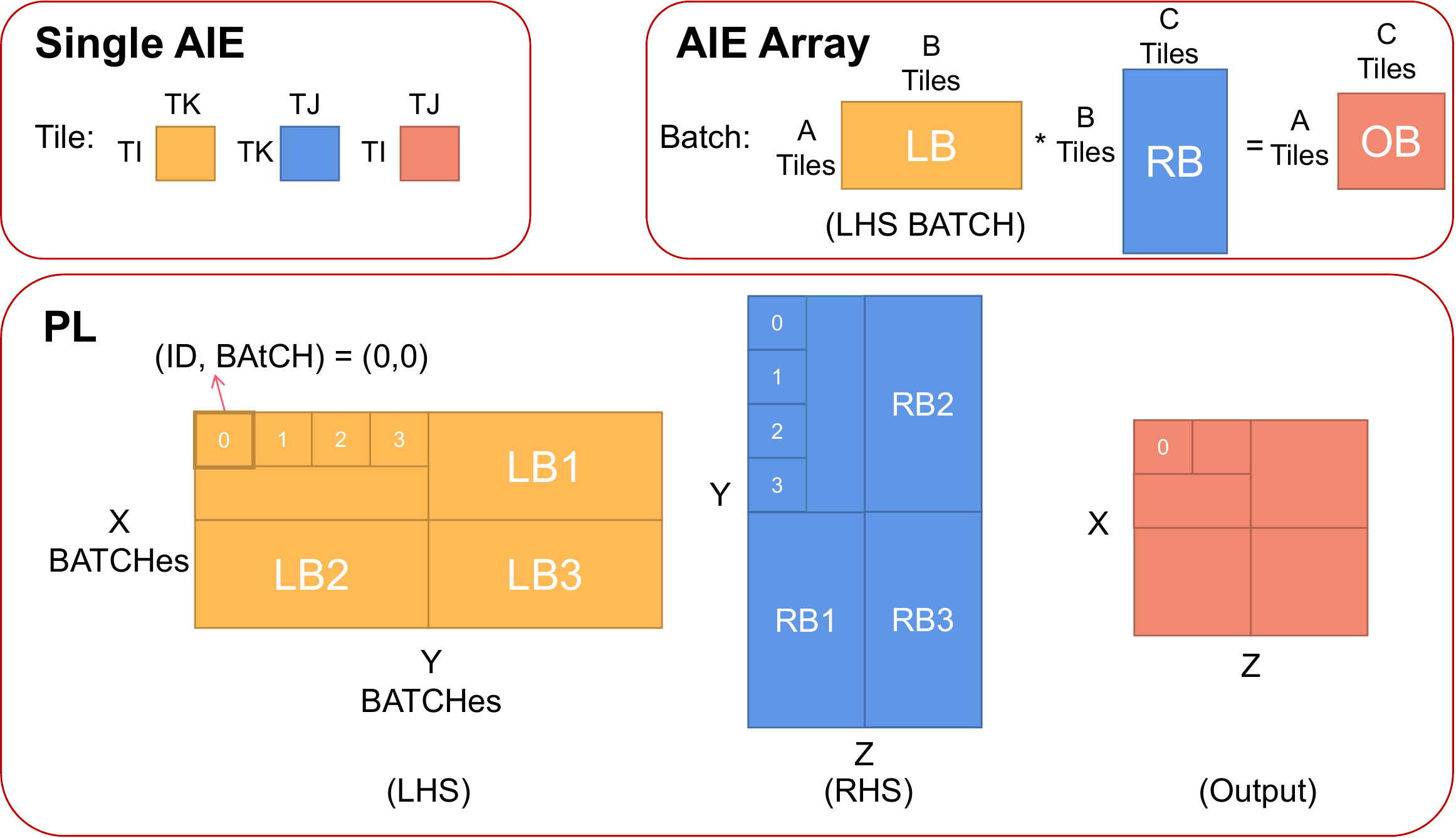}
\caption{\textcolor{black}{Mapping strategy and data layout.}}
\label{fig:Mapping Strategy and Data Layout}
\vspace{-10pt}
\end{figure}

Four levels of tiling and output stationary dataflow are applied in our design to compute the matrix-matrix multiply(MM). The pseudo-code and the corresponding mapping strategy of our tiled MM example are shown in Listing~\ref{code:Data flow} with four levels of loops and Fig~\ref{fig:Mapping Strategy and Data Layout} respectively.


\begin{figure}
\begin{lstlisting}[language=python,label=code:Data flow ,caption=MM loop tiling and dataflow.]
# Sequential loop: from off-chip to on-chip
for m.0 in range(M/(TI*A*X)):
for n.0 in range(N/(TJ*C*Z)):
for k.0 in range(K/(TK*B*Y)):
   dataMovementOffChip2OnChip(...)
# Sequential loop: reuse PL on-chip buffer
   for m.1 in range(X):
   for n.1 in range(Z):
   for k.1 in range(Y):
      dataMovementOnChip2AIE(...)
# Parallel loop: AIE Array
      for m.2 in range(A):
      for n.2 in range(C):
      for k.2 in range(B):
# Single AIE loops with 2D-SIMD Instructions
         for m.3 in range(TI/PI):
         for n.3 in range(TJ/PJ):
         for k.3 in range(TK/PK):
            Matmul(m.3, n.3, k.3)
\end{lstlisting}
\vspace{-10pt}
\end{figure}
\noindent\textbf{Single AIE Level (Line 15-19).} An MM with size TI * TK * TJ named ``TILE" is mapped to a single AIE. To fully utilize the 7-way VLIW capability of the AIE core,  We manually pack several 2D-SIMD vector intrinsics into a function ``MatMul" to calculate a sub-tile with size PI * PK * PJ. Thus a TILE can be computed by launching "MatMul" (TI/PI) * (TJ/PJ) * (TK/PK) times.

\noindent\textbf{AIE Array Level (Line 11-14).} When scaling out to the AIE array, we explore the spatial data parallelism among different AIEs as shown in the AIE array mapping in Fig~\ref{fig:Mapping Strategy and Data Layout}. More specifically, we unroll A * B * C TILEs on the AIE array with each AIE computing a TILE as mentioned above. The TILEs in the same reduction dimension (k.2 loop) are assigned to the AIEs in the same column producing the read-after-write (RAW) dependency. The m.2 and n.2 loop are mapped to different columns in the AIE array. We refer to the MM with size (A*TI) * (B*TK) * (C*TJ) as the ``BATCH'' level.

\noindent\textbf{PL On-chip Data Reuse Level (Line 6-10).} In order to amortize the bandwidth gap between off-chip to PL and PL to AIE array, we explore the on-chip data reuse by allocating a large number of RAMs on the PL side to store multiple X * Y * Z BATCHes of data. The BATCHes of data are fed to the AIE array by the DMA module on the PL side following the bubble-free pipeline algorithm which will be discussed in section~\ref{sec:bubble-free scheduling} and the partial result from the AIE array will finally be accumulated on the PL side.

\noindent\textbf{Off-chip Level (Line 1-5).} Data that exceeds the capacity of the on-chip buffer are stored in the off-chip memory. The double buffer technique is applied to hide the overhead of loading/storing the data between off-chip to on-chip memory.

\subsection{Single AIE Programming Model}
Our system-level design starts from the single AIE kernel. The Vitis programming tools expose C intrinsics ~\cite{aie2021Intrinsics}, including load/store, scalar, and vector operations, for AIE programming. To achieve high computation efficiency of AIE, it is necessary for us to explore the best coding style for a single AIE. 
\label{sec:single_aie}

\begin{figure}
\begin{lstlisting}[language=C, label=code:Single_kernel,caption=High efficiency single kernel coding style for matrix multiplication.]
#define PI 8
#define PJ 2
#define PK 8
void mm_kernel(
  input_window_float * restrict L,// LHS
  input_window_float * restrict R, // RHS
  output_window_float * restrict O ) { // Output
  preload(L,R);  //Load data from local mem to reg
  for(int m.3 = 0; m.3 < TI/PI; m.3++){
  chess_pipelining  // Apply software pipelining
    for(int n.3 = 0; n.3 < TJ/PJ; n.3++){
      v8float acc0 = null_v8float(); //Set Acc reg
      v8float acc1 = null_v8float(); //to zero
      for(int k.3 = 0; k.3 < TK/PK - 1; k.3++){
        MatMul_without_store([acc0; acc1],
        L(m.3, k.3), R(k.3, n.3) );}
      MatMul_with_store([acc0; acc1],
      L(m.3,TK/PK-1),R(TK/PK-1,n.3),O(m.3, n.3));}}}
      //Hoist the final loop to store data from 
      //reg to local mem
\end{lstlisting}
\vspace{-25pt}
\end{figure}

\begin{figure}
\centering
\includegraphics[width=1\linewidth]{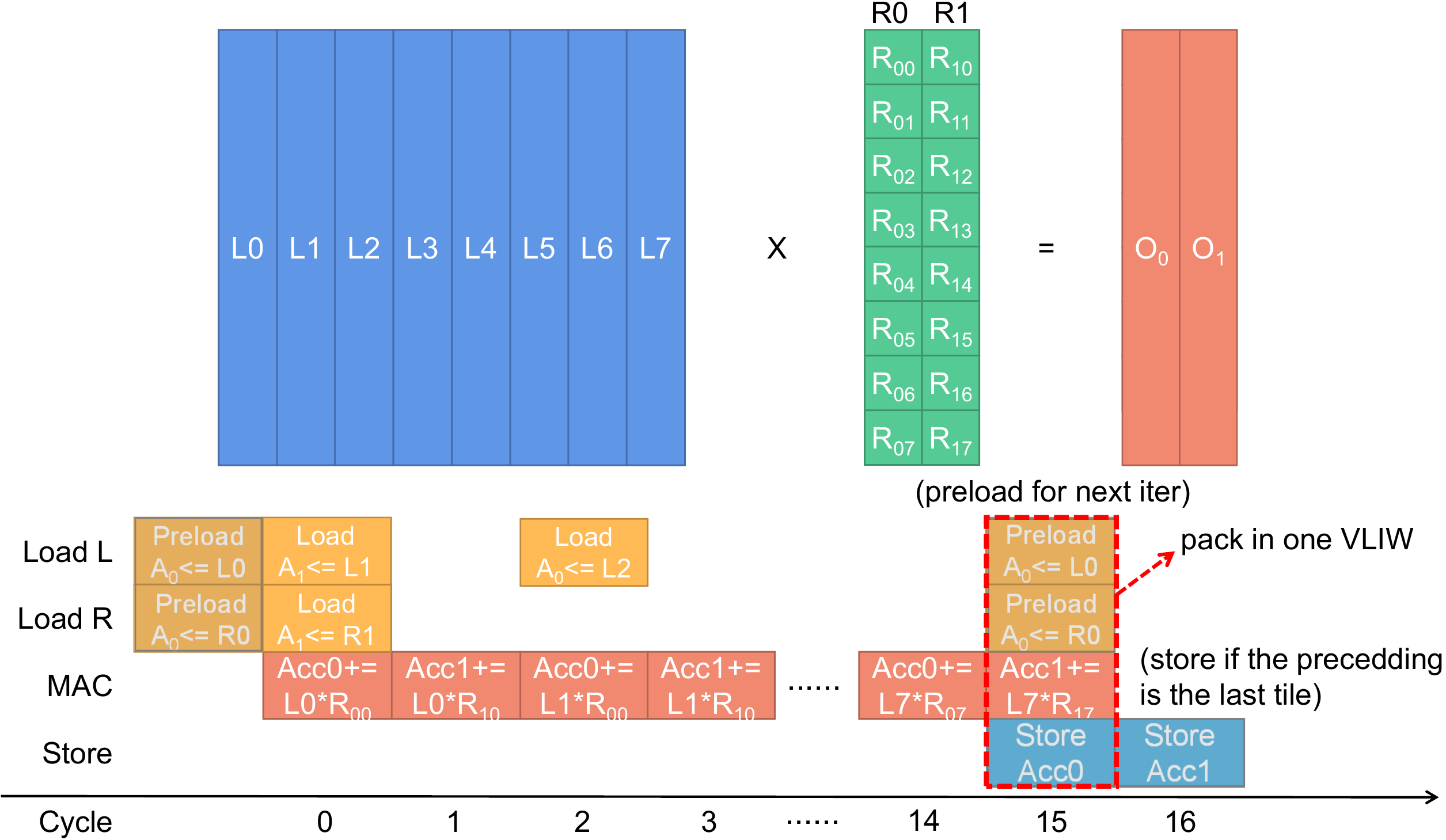}
\caption{Single AIE pipeline.}
\label{fig:single aie pipeline}
\vspace{-18pt}
\end{figure}

The overall data processing in a single AIE is shown in Listing~\ref{code:Single_kernel}. Variables \texttt{L}, \texttt{R}, and \texttt{O} are three pointers referencing the local memories allocated for the MM kernel(Lines 5-7).  \texttt{restrict} directives specify that input pointers do not alias, enabling more aggressive optimizations. \texttt{chess\_pipelining} is applied for all the three loops(Line 10) to inform the compiler of finding optimized pipeline design. To reduce the frequency of writing local memory \texttt{O}, we choose \texttt{k} loop as the innermost loop (Line 14) and introduce two 8-length vector registers, \texttt{acc0} and \texttt{acc1} (line 12-13), to hold the partial accumulation results in an interleaved manner which avoids of waiting for two cycles adding the partial result to the same register after each vector MAC operation.  This allows the local memory \texttt{O} to be written only once after the final accumulation results are carried out. To make full usage of the upto 7-way VLIW and get back-to-back issued MAC operations, we manually pack 16 8*1 vector 2D-SIMD instructions in each Matmul function to calculate MM with the size of PI(8)*PK(8)*PJ(2) (Line 1-3). In addition to two accumulator registers, we further allocate four 8-length in total 1Kb vector registers (A0, A1, B0, B1) shown in Fig.~\ref{fig:single aie pipeline} for storing the two vector operands needed for current MAC operation and two pre-loaded vector operands for future MAC operation. We use Li and Ri to notify the 8-length vector and Rij to notify the element in one vector. By pre-loading L0 and R0 from local memory to vector register A0 and B0 prior to the start of the Matmul function (Line 8), the MAC instruction can be issued in time 0. And at the same time, the two load instructions for loading the local memory used in future MAC operations can be packed to the same VLIW. Since only when the last iteration should we store data from accumulator register Acc0 and Acc1 back to local memory so there are two kinds of Maltul functions in the design(Lines 15 and 18). Note that we hoist the last iteration of the loop out to avoid the  significant performance degradation of inserting an \texttt{if} statement in the \texttt{k.3} loop.

In summary, to conduct MM under FP32 datatype on a single AIE we pack 16 MAC8 together in the innermost loop as an atomic operation and these 16 instructions will calculate a Matmul block with size 8*8*2. 
To scale up the MM size, we can assign TI, TJ, and TK sizes that are multiple of our atomic operation, for example, 32x32x32. 
In this case, the loop boundaries for Line 9, Line 11, and Line 14 are 4, 16, and 3 respectively. 
The methodologies of building atomic and scaling up MM size in a single AIE are applied to other data types as well.

\subsection{Scaling Out to AIE Array}
\begin{figure}
    \centering
    \includegraphics[width=0.8\linewidth]{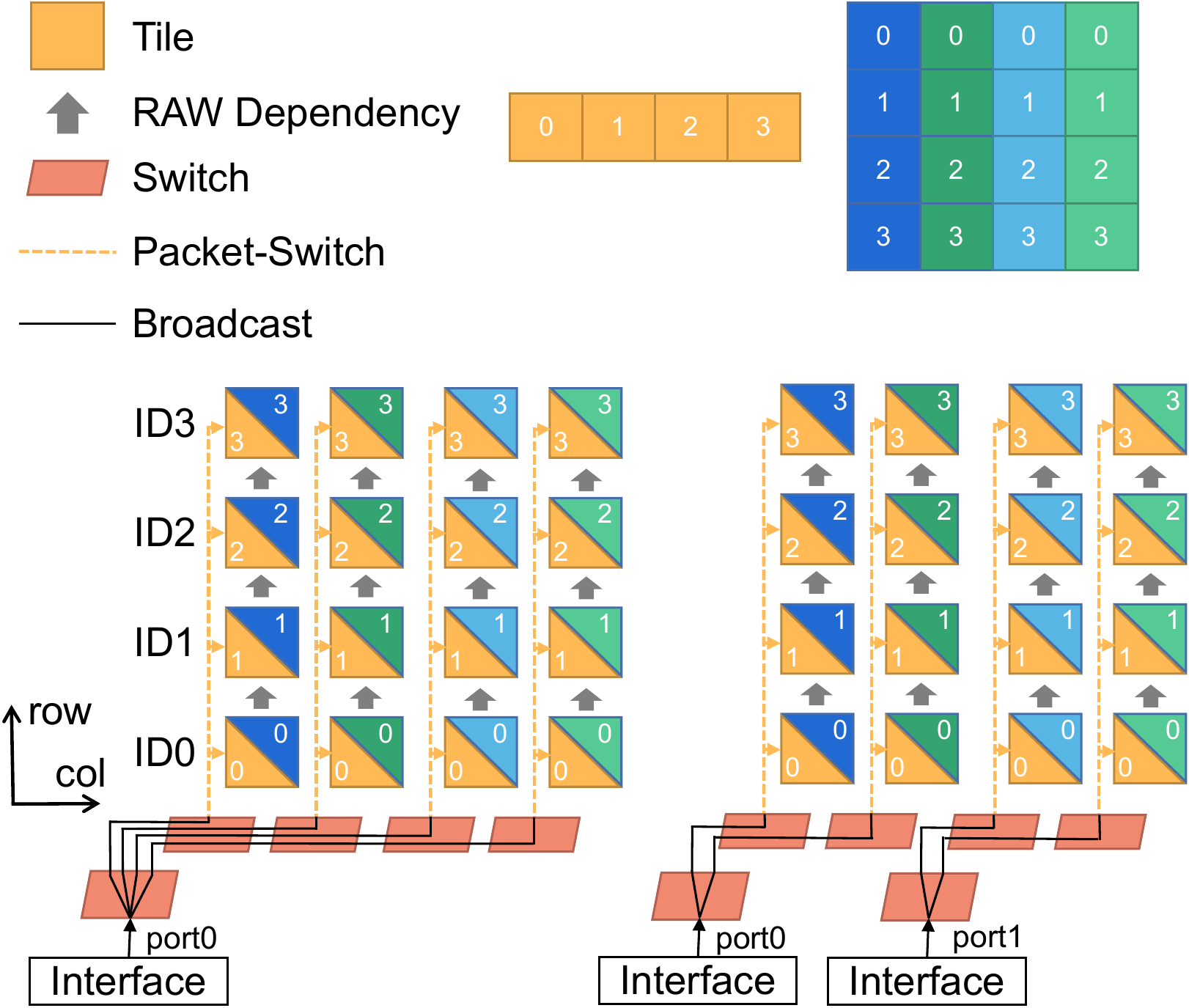}
    \caption{Combining broadcast circuit-switched and packet-switched connections to reduce required I/Os to AIE array.}
    \label{fig:AIE Mapping and IO design}
\end{figure}

\noindent\textbf{PLIO Reuse.}
When scaling out to a large number of AIE cores, as described in section \ref{sec:3_Versal_arch}, the total number of PLIOs in the interface tiles is much smaller than the total number of operands of all the AIE cores, identifying reuse patterns within the AIE array can be important to build a feasible and communicating and computation balanced AIE array design. 
As shown in Fig~\ref{fig:AIE Mapping and IO design} the 4x4 AIE array calculates MM with 1*4*4 TILEs in which the AIEs in the same column take the output of the previous AIE as input producing RAW dependency. Figure~\ref{fig:AIE Mapping and IO design} (a) demonstrates how the 1*4 TILEs in LHS matrix is transferred to 16 AIEs by reusing one port in the interface tile. A similar mechanism can be applied to the RHS and Output matrix. 
In particular, we leverage a combination of broadcast and packet-switch connections to effectively transfer the data from PL to AIE throughput the I/O port in interface tiles. First, by using the data broadcast opportunities in the MM application (e.g. one row of TILEs in LHS can be broadcast to different columns of RHS), we can use 1 port to broadcast the single TILE(0,0) of LHS to AIE(col 0-3, row 0) as shown in solid lines. 
The packet-switch opportunity appears when the computation time of a single AIE is higher than communication, i.e., the CTC ratio of a single AIE is larger than 1. In this case, by attaching the different data TILEs with a unique header, the data TILEs can be scattered to multiple AIEs in a time-division multiplex way without hurting the computation of each AIE.
For example, a single AIE kernel that computes 32x32x32 MM with FP32 data type takes at least 4096 cycles to compute and 1024 cycles to transfer LHS and RHS TILEs. 
In this case, the single AIE kernel CTC ratio is 4. Here we refer to 1024 cycles as one time step. 
Therefore, we can pack 4 LHS TILEs (0, 0-3)(same for RHS) in the same packet stream to AIE(col 0, row 0-3) on different time steps as shown in the dashed lines.
In a summary, TILE 0 of LHS can be broadcast to AIE(col 0-3, row 0) in time step 0, TILE 1 of LHS can be broadcast to AIE(col 0-3, row 1) in time step 1 by reusing the same port. TILE 2 and 3 of LHS share the same pattern in time steps 2 and 3.
Thus, by combining broadcast circuit-switched connections and packet-switched connections, we can use 1 port to distribute data to 16 AIEs in four time steps without performance degradation which reduces the number of ports by 16x.

\noindent\textbf{Routing Optimization.} By combining the broadcast and packet-switch connections we hugely reduce the ports needed in the design, however, the routing complexity is not reduced for each switch box. Currently, we observe that the Vitis AIE compiler will split the data stream immediately in the first switch box after the interface tile as shown in ~\ref{fig:AIE Mapping and IO design} (a). Thus, routing congestion in the switch boxes is very likely to happen when broadcasting data to AIEs at a long distance from the interface tile. In order to reduce the routing congestion caused by long-distance broadcasts, we apply broadcast factors on both LHS and RHS matrices. As shown in Fig. ~\ref{fig:AIE Mapping and IO design} (b), instead of broadcasting the LHS to all four columns of the AIE array, we set the broadcast factor to two which means that we use 2 ports with each one sending the same data to two columns. Thus the total number of connections from west to east is reduced from 10 to 4. The benefit will be more obvious when routing on more AIEs.

\subsection{AIE-PL Bubble-free Pipelining Data Transfer Algorithm}
\label{sec:bubble-free scheduling}
\begin{figure}
    \centering
    \includegraphics[width=1\linewidth]{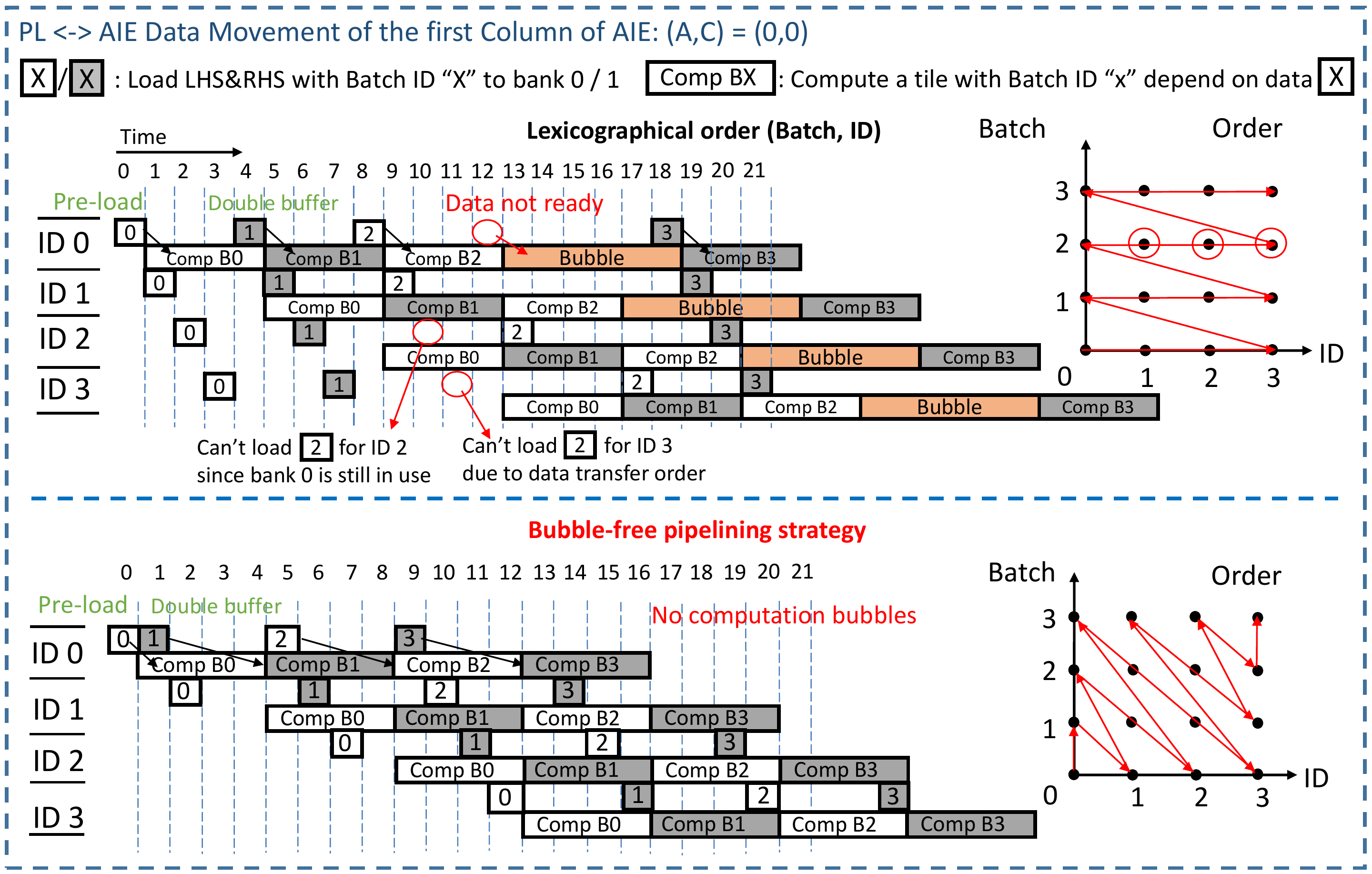}
    \caption{\textcolor{black}{Bubble free data movement between PL and AIE.}}
    \label{fig:Bubble_free_strategy}
    \vspace{-20pt}
\end{figure}

In order to amortize the bandwidth gap between off-chip memory to PL and PL to AIE, we hugely explore the on-chip data reuse by allocating over 80\% on-chip buffers and storing multiple numbers of BATCHes. We design dedicated DMA modules with a bubble-free pipelining algorithm that determines the order of each TILE that reaches the corresponding local memory of AIEs. We use the data movement and computation in AIE column 0, namely AIE(col 0, row 0-3) with ID0-ID3 that calculates the first row of LHS and column of RHS in BATCH 0-3 shown in Fig~\ref{fig:Mapping Strategy and Data Layout}, as an example to demonstrate our data transferring strategy. In Figure~\ref{fig:Bubble_free_strategy}, we first illustrate the pipeline bubbles when using the straightforward data transferring sequence where multiple BATCHes of data are sent to the AIE array in the lexicographical order as (BATCH, ID). Lexicographical order means that the TILE with the smaller BATCH index will be transferred earlier than the larger BATCH index. It also means that the TILE with a smaller TILE ID in the same BATCH will be transferred earlier than the larger TILE ID. As demonstrated in Figure~\ref{fig:Bubble_free_strategy}, each TILE has a unique (BATCH, ID) pair and we use white or grey to identify loading the LHS and RHS data into the ping-pong banks of each AIE local memory. The time for storing the data in local memory is overlapped by the computation due to VLIW, thus omitted in the figure. Once the previous AIE finishes computing, the read-after-write (RAW) dependency between AIEs in a column is considered resolved. For illustration purposes, We assume the CTC ratio of each AIE is 4, which means 1 time step for data loading and 4 time steps for computation. The order graph on the right side of Figure~\ref{fig:Bubble_free_strategy} illustrates the sequence of (BATCH,ID) during data transferring. When applying the lexicographical order, from time 0 to time 3, ID 0 to ID 3 in BATCH 0 are transferred. From time 4 to time 7, ID 0 to ID 3 in BATCH 1 are transferred. If there are no bubbles, from time 8 to time 11, ID 0 to ID 3 in BATCH 2 will be transferred. However, the first data transfer bubble appears in time 10 for AIE 2. AIE 2 takes the BATCH 0 data in time 2 and BATCH 1 data in time 6. It does not compute on the BATCH 0 data until time 9 as the initial latency due to RAW dependencies from AIE 1 and AIE 0. It is impossible for AIE 2 to take BATCH 2 data until it completes the execution of BATCH 0 and releases the memory bank 0. Therefore, it causes three transferring bubbles and pushes back the BATCH 2 data transfer from time 10 to time 13. Then butterfly effect happens due to the lexicographical order, AIE 0 can't get its BATCH 3 data, thus after finishing computing BATCH 2, it can not start to compute BATCH 3 which leads to computation bubbles from time 13-18. 

To address this, we implement a pipeline bubble-free scheduling technique as shown in Fig.~\ref{fig:Bubble_free_strategy}. In this approach, we only send data that is needed in the next computation period. For example, in the first computation period, corresponding to time 1-4, we send data with (BATCH, ID) pair (1,0) and (0,1) sequentially. These two tiles are needed in time 5-8 for AIE 0 and AIE 1. Similarly, three tiles are sent in time 5-7 as they are needed in time 9-12 for AIE 0, 1, and 2. By using this zigzag data transferring manner instead of lexicographical order between PL and AIE, we successfully eliminate data transfer bubbles \& compute bubbles and achieve a full pipeline.

\subsection{AutoMM Framework}
The overall architecture of our AutoMM framework is shown in Fig~\ref{fig:framework}. Our AutoMM framework mainly consists of four components including the Python-based user interface, the optimizer for design space exploration(DSE), the host CPU Runtime manager and the automatic code generator(ACG). We provide users with Python APIs shown in Listing~\ref{code:automm_mlp} that take the definition of the MM-based model as input (Lines 2-6). Based on the size of the MM kernels, our DSE (Lines 8-9) will find the best hardware configuration by optimizing the 
 AIE PLIO utilization, AIE placement, PL buffer utilization, DMA between AIE $\leftrightarrow$ PL, and the overall tiling for both PL and AIE. Then our ACG (Lines 10-11) takes the output of DSE and Host Runtime to generate the corresponding source code for AIE array, PL and Host CPU respectively. Our API also enables automatic compilation via AMD Vitis compiler (Lines 13-14) and on-board execution (Lines 15-16). To the best of our knowledge, AutoMM is the first work to provide the high-level Python APIs to generate source code for Versal ACAP.


\begin{lstlisting}[language=Python, label=code:automm_mlp,caption=AutoMM Python APIs.]
import charm
#Define the LHS(A) and RHS(B) operands
A=np.random.rand(4096, 4096).astype(np.float32)
B=np.random.rand(4096, 4096).astype(np.float32)

automm=charm() #Create charm object

#Launch CHARM dse
Versal_config=automm.cdse(A,B)
#Launch CHARM Code Gen
automm.cacg(Versal_config,'vck190')

#Run Vitis Compilation Flow
automm.build()
#Run On-board Execution
automm.run()
\end{lstlisting}

\vspace{-10pt}
\section{Experiment Results}
\label{sec:5_Experiment}

In this section, we demonstrate the MM design performance, power, and energy efficiency of {\dac} implementation on AMD VCK190 under various data types. 
We compare them with prior works on other platforms including AutoSA~\cite{wang2021autosa} implementation on AMD U250 FPGA, cuBLAS~\cite{cublas} on Nvidia A100 40GB PCIe GPU and Jetson TX2 GPU. 
We also evaluate {\dac} on two deep learning inference tasks: NCF~\cite{he2017neural} for recommendations, MLP~\cite{wang2019benchmarking} for multilayer perceptron classification or regression. These two inference models are mainly based on different shapes of matrix-multiply layers.

\vspace{-5pt}
\subsection{Experiment Setup}
AMD Vitis 2021.1 is used for all the experiments on VCK190 with PL running on 230MHz and AIE running on 1GHz. The designs on U250 FPGA are generated by AutoSA~\cite{wang2021autosa} and Autobridge~\cite{guo2021autobridge} for FP32, INT8 (300MHz) and INT16 (250MHz) using AMD Vitis 2019.2.

We set up the GPU experiment of MM under FP32 data type by using cublasSgemm() API in cuBLAS from CUDA Toolkit 10.2 for Jetson TX2 GPU and 11.3 for A100 GPU. For INT8 experiment on A100 GPU, we use the cublasGemmEx() API in cuBLAS from CUDA 11.3.

When comparing the performance of MM, we use the same size for VCK190 and NVIDIA GPUs. 
For U250 designs, we pick the design sizes with the best performance due to the AutoSA~\cite{wang2021autosa} design size limitation. 
We set the matrix size to 6K*6K*6K for VCK190, Nvidia A100, and Jetson TX2 GPUs, and 1040*1K*1K for U250 under FP32.  
For INT16, the matrix size is 9K*9K*10K and 1K*1K*1K for VCK190 and U250. 
For INT8, the matrix size is 16K*16K*16K for VCK190 and Nvidia A100 GPU, and 1056* 1K*1K for U250.

\begin{table}
\Huge
\caption{Performance, power, and energy efficiency comparisons among FPGAs and GPUs when the data type is INT16. AMD VCK190 achieves gains of 3.26x than AMD U250. GPUs Jetson TX2 and A100 do not support INT16.}
\vspace{-10pt}
\label{tbl:comparison_energy_INT16}
    \begin{center}
    \begin{adjustbox}{width=1\columnwidth,center}
\begin{tabular}{ c | c | c | c | c c }

 \toprule
  \multirow{2}{*}{\textbf{\textcolor{black}{Fabrication}}}  
  &  \multirow{2}{*}{\textbf{\textcolor{black}{Board Name \& Platform}}}
  &  \textbf{\textcolor{black}{Performance}} 
  & \textbf{\textcolor{black}{Power}} 
  & \multicolumn{2}{c}{\textbf{\textcolor{black}{Energy Efficiency}}} 
\\ & & (GOPS) & (Watt) & (GOP/J) & (Ratio) \\
    \midrule
    \multirow{2}{*}{16 nm} 
                        & \multicolumn{1}{c}{AMD U250~\cite{u250_web}, AutoSA~\cite{wang2021autosa}}  
                        & \multicolumn{1}{c}{3,450}
                        & \multicolumn{1}{c}{85.02} 
                        & \multicolumn{1}{c}{40.58}
                        & \multicolumn{1}{c}{1.00x} 
                        \\
                        & \multicolumn{1}{c}{Nvidia Jetson TX2~\cite{tx2_web}, cuBLAS~\cite{cublas}}  
                        & \multicolumn{1}{c}{N/A}
                        & \multicolumn{1}{c}{N/A} 
                        & \multicolumn{1}{c}{N/A}
                        & \multicolumn{1}{c}{N/A} 
                        \\
                        \hline
    \multirow{2}{*}{7 nm}
                        & \multicolumn{1}{c}{\cb{AMD VCK190~\cite{vck190_web}, \textbf{This work}}} 
                        & \multicolumn{1}{c}{\cb{7,511}}
                        & \multicolumn{1}{c}{\cb{56.82}} 
                        & \multicolumn{1}{c}{\cb{132.20}}
                        & \multicolumn{1}{c}{\cb{3.26x}} 
                        \\
                        & \multicolumn{1}{c}{Nvidia A100~\cite{a100_web}, cuBLAS~\cite{cublas}}
                        & \multicolumn{1}{c}{N/A}
                        & \multicolumn{1}{c}{N/A} 
                        & \multicolumn{1}{c}{N/A}
                        & \multicolumn{1}{c}{N/A} 
                        \\

    \bottomrule
\end{tabular}
\end{adjustbox}
    \end{center}
\vspace{-10pt}
\end{table}
\begin{table}
\Huge
\caption{Performance, power, and energy efficiency comparisons among FPGAs and GPUs when the data type is INT8. AMD VCK190 achieves gains of 1.70x energy efficiency than Nvidia A100, 6.23x than AMD U250.}
\vspace{-10pt}
\label{tbl:comparison_energy_INT8}
    \begin{center}
    \begin{adjustbox}{width=1\columnwidth,center}
\begin{tabular}{ c | c | c | c | c c }

 \toprule
  \multirow{2}{*}{\textbf{\textcolor{black}{Fabrication}}}  
  &  \multirow{2}{*}{\textbf{\textcolor{black}{Board Name \& Framework}}} 
  &  \textbf{\textcolor{black}{Performance}}  
  & \textbf{\textcolor{black}{Power}} 
  & \multicolumn{2}{c}{\textbf{\textcolor{black}{Energy Efficiency}}} 
\\ & & (GOPS) & (Watt) & (GOP/J) & (Ratio) \\

    \midrule
    \multirow{2}{*}{16 nm} 
                        & \multicolumn{1}{c}{AMD U250~\cite{u250_web}, AutoSA~\cite{wang2021autosa}} 
                        & \multicolumn{1}{c}{6,740}
                        & \multicolumn{1}{c}{90.90} 
                        & \multicolumn{1}{c}{74.15}
                        & \multicolumn{1}{c}{1.00x} 
                        \\
                        & \multicolumn{1}{c}{Nvidia Jetson TX2~\cite{tx2_web}, cuBLAS~\cite{cublas}} 
                        & \multicolumn{1}{c}{N/A}
                        & \multicolumn{1}{c}{N/A} 
                        & \multicolumn{1}{c}{N/A}
                        & \multicolumn{1}{c}{N/A} 
                        \\
                        \hline
    \multirow{2}{*}{7 nm}
                        & \multicolumn{1}{c}{\cb{AMD VCK190~\cite{vck190_web}, \textbf{This work}}}
                        & \multicolumn{1}{c}{\cb{28,150}}
                        & \multicolumn{1}{c}{\cb{60.96}} 
                        & \multicolumn{1}{c}{\cb{461.74}}
                        & \multicolumn{1}{c}{\cb{6.23x}} 
                        \\
                        & \multicolumn{1}{c}{Nvidia A100~\cite{a100_web}, cuBLAS~\cite{cublas}}
                        & \multicolumn{1}{c}{67,200}
                        & \multicolumn{1}{c}{248.08} 
                        & \multicolumn{1}{c}{270.88}
                        & \multicolumn{1}{c}{3.65x}
                        \\

    \bottomrule
\end{tabular}
\end{adjustbox}
    \end{center}
    \vspace{-10pt}
\end{table}
   

\begin{table}[!tb]
\Huge
\caption{Resource utilization of MM Acc on VCK190.}
\vspace{-10pt}
\label{tbl:Hardware Utilization}
\begin{center}
\begin{adjustbox}{width=1\columnwidth,center}
\begin{tabular}{c | c | c | c | c | c | c | c}
   
\toprule
\textbf{Datatype}  & \textbf{REG} &  \textbf{LUTLogic} &  \textbf{LUTMem} &  \textbf{BRAM} & \textbf{URAM}  & \textbf{DSP} & \textbf{AIE}\\ 
\midrule
INT8   & 91185 (5.18\%)  &  84072 (9.53\%)  &   1001   (0.22\%)  & 669 (69.18\%)  & 384 (82.94\%) & 71(3.61\%) & 192 (48\%)\\ 
INT16 & 126773 (7.23\%)  &  91664 (10.44\%)  &   999   (0.23\%)  & 477 (49.33\%)  & 384 (82.94\%) & 93(4.73\%) & 288(72\%)\\
FP32 & 87790 (5.00\%)  &  63845 (7.24\%)  &   1004   (0.23\%)  & 661 (68.36\%)  & 384 (82.94\%) & 163(8.28\%) & 384(96\%) \\
\bottomrule
\end{tabular}
\end{adjustbox}
\end{center}
\vspace{-20pt}
\end{table}

We use AMD board evaluation and management tool~\cite{BEAM}, AMD AMD Board Utility~\cite{Xbutil}, NVIDIA System Management Interface tool, and P3 P4460 Kill-A-Watt(Tm) power meter to measure the power of VCK190, U250 FPGA, A100, and Jetson TX2 GPU respectively. 
We iterate the design to make sure the total execution time exceeds the 60s and the power is relatively stable and the average value is reported.



\subsection{Comparison with Prior FPGA and GPUs}

In this section, we compare our design with prior FPGA and GPU work under FP32, INT16, and INT8 data types demonstrated in Table~\ref{tbl:comparison_energy_efficiency}, Table~\ref{tbl:comparison_energy_INT16} and Table~\ref{tbl:comparison_energy_INT8} respectively. The hardware resource utilization is shown in Table~\ref{tbl:Hardware Utilization}.  {\dac} achieves 3.7 TFLOPs on FP32 data type by using 384 AIEs. For INT16 design, since the routing congestion becomes the bottleneck preventing us from using more AIEs, {\dac} achieves 7.5 TOPs throughput using 288 AIEs. 
The computation capacity of a single AIE for the INT8 data type is 128x of the FP32 data type.
The CTC ratio for the INT8 is half of the FP32 and the INT8 AIE array design is bounded by the number of PLIO. 
By using 192 AIEs, {\dac} achieves 28.2 TOPs on the INT8 data type. 
We compare the throughput and energy efficiency of the MM application with the state-of-the-art polyhedral-based framework AutoSA on prior generation U250 FPGA under FP32, INT16, and INT8 data types. {\dac} achieves 7.20x, 3.26x and 6.23x energy efficiency respectively. 
We also compare the throughput and energy efficiency of two Nvidia GPUs. {\dac} achieves 2.32x higher energy efficiency than Nvidia Jetson TX2 under FP32, and 1.06x, 1.70x higher energy efficiency than Nvidia A100 under FP32 and INT8 respectively.



\subsection{End-to-end Applications}
\begin{table}
\Huge
\caption{Energy efficiency comparisons of GPU A100 PyTorch and VCK190 {\dac} for two FP32 end-to-end deep learning inference applications: NCF \& MLP.}
\vspace{-10pt}
\label{tbl:app_compare}
\begin{center}
\begin{adjustbox}{width=1\columnwidth,center}
\begin{tabular}{c | c c c c c c}
   
\toprule
\multirow{3}{*}{\textbf{Application}}  & \multicolumn{3}{c}{\textbf{GPU A100 PyTorch}} & \multicolumn{3}{c}{\textbf{AMD VCK190 {\dac}}} \\
    & Performance & Power & Energy Eff. & Performance & Power & Energy Eff.\\
    & (GFLOPS) & (Watt) & (Ratio) & (GFLOPS) & (Watt) & (Ratio)\\
\midrule
NCF~\cite{he2017neural} & 12,801.37 & 248.53 & 1.00x & \cb{2,265.09} & \cb{45.85} & \cb{0.96x} \\
MLP~\cite{wang2019benchmarking} & 13,668.87 & 248.32 & 1.00x & \cb{3,473.86} & \cb{54.33} & \cb{1.16x} \\
\bottomrule
\end{tabular}
\end{adjustbox}
\end{center}
\vspace{-20pt}
\end{table}

We apply our {\dac} framework to NCF and MLP applications and compare the energy efficiency with A100 GPU under FP32 data type. {\dac} achieves 2.3 TFLOPs and 0.96x energy efficiency compared with A100 GPU shown in Table~\ref{tbl:app_compare} since the MM with small sizes in NCF leads to performance degradation on the overall performance. For MLP, {\dac} achieves 3.5 TFLOPs and 1.16x energy efficiency gain compared with A100 GPU.


\section{Conclusion and Acknowledgement}
\label{sec:6_Conclusion}
\vspace{-5pt}
In this work, we propose {\dac} framework, an automatic white-box tool that can systematically generate the design for MM accelerators under different data types on Versal. 
We believe our design methodology can be a good reference for other users to design their own applications on Versal.

We thank all the reviewers for their valuable feedback.
We acknowledge the support from the University of Pittsburgh New Faculty Start-up Grant, Pitt Center for Advanced Manufacturing (UPCAM) Grant, Pitt Provost Open Educational Resources (OER) Grant, NSF awards CNS-2213701, CCF-2217003.
We thank AMD for FPGA and software donation, the AMD Heterogeneous Accelerated Compute Cluster at UCLA, and the Center for Research Computing (CRC) at the University of Pittsburgh.

\footnotesize{
\bibliographystyle{IEEEtran}
\balance
\bibliography{reference}
}

\end{document}